\documentclass[structabstract]{aa}  
\usepackage{graphicx}
\usepackage{txfonts}

\usepackage{natbib}
\usepackage{enumerate}

\bibpunct{(}{)}{;}{a}{}{,} 

\begin{document}

\newcommand{\xco}{$X_{\mathrm{CO}\,}$}

   \title{The Herschel Virgo Cluster Survey. IX. Dust-to-gas mass ratio and metallicity gradients in four Virgo spiral galaxies}

\author{Laura Magrini\inst{1}, 
Simone Bianchi\inst{1},   
Edvige Corbelli\inst{1},
Luca Cortese\inst{3},
Leslie Hunt\inst{1},
Matthew Smith\inst{2},
Catherine Vlahakis\inst{4},
Jonathan Davies\inst{2},
George J. Bendo\inst{5}, 
Maarten Baes\inst{6},
Alessandro Boselli\inst{7},
Marcel Clemens\inst{8},
Viviana Casasola\inst{9},
Ilse De Looze\inst{6},
Jacopo Fritz\inst{6},
Carlo Giovanardi\inst{1},
Marco Grossi\inst{10},
Thomas Hughes\inst{11}, 
Suzanne Madden\inst{12},
Ciro Pappalardo\inst{1},
Michael Pohlen\inst{2},
Sperello di Serego Alighieri\inst{1},
Joris Verstappen\inst{6}
}

\institute{
INAF--Osservatorio Astrofisico di Arcetri, Largo E. Fermi, 5, I-50125 Firenze, Italy
\email{laura@arcetri.astro.it}
\and
Department of Physics and Astronomy, Cardiff University, The Parade, Cardiff, CF24 3AA, UK 
\and
European Southern Observatory, Karl-Schwarzschild-Strasse 2, D-85748 Garching bei Mnchen, Germany 
\and
Departamento de Astronomia, Universidad de Chile, Casilla 36-D, Santiago, Chile
\and
Jodrell Bank Centre for Astrophysics, Alan Turing Building, School of Physics and Astronomy, University of Manchester, Manchester M13 9PL 
\and
Sterrenkundig Observatorium, Universiteit Gent, Krijgslaan 281 S9, B-9000 Gent, Belgium 
\and
Laboratoire d'Astrophysique de Marseille, UMR 6110 CNRS, 38 rue F. Joliot-Curie, F-13388 Marseille, France
\and
Osservatorio Astronomico di Padova, Vicolo dell'Osservatorio 5, 35122 Padova, Italy
\and
INAF -  Istituto di Radioastronomia, via P. Gobetti 101, 40129 Bologna, Italy 
\and
CAAUL, Observatorio Astronomico de Lisboa, Universidade de Lisboa, Tapada de Ajuda, 1349-018, Lisboa, Portugal 
\and
The Kavli Institute for Astronomy \& Astrophysics, Peking
University, Beijing 100871, China
\and
Service d'Astrophysique, CEA/Saclay, l'Orme des Merisiers, 91191 Gif-sur-Yvette, France 
 \\ 
}

   \date{Received ; accepted }

 
  \abstract
{Using Herschel data from the Open Time Key Project  {\em the Herschel Virgo 
Cluster Survey (HeViCS)}, 
we investigated the relationship between the metallicity gradients expressed 
by metal abundances in the gas phase as traced by the chemical composition of HII regions, 
and in the solid phase, as traced by the dust-to-gas mass ratio.  }
{We derived the radial gradient of the dust-to-gas mass ratio for all 
galaxies observed by HeViCS whose  metallicity gradients are available in the literature. 
They are all late type Sbc galaxies, namely NGC4254, NGC4303, NGC4321, and NGC4501. }
{We fitted PACS and SPIRE observations with a single-temperature modified
blackbody, inferred the dust mass, and 
calculated two dimensional maps of the dust-to-gas mass ratio, 
with the total mass of gas from available HI and CO  maps. 
HI moment-1 maps were used to derive the geometric parameters of the 
galaxies and extract the radial profiles.   
We examined  different dependencies on metallicity of the CO-to-H$_2$ conversion  factor 
(\xco), used to  transform the $^{12}$CO observations into the amount of molecular hydrogen. }
{We found that in these galaxies the  dust-to-gas mass ratio radial profile 
is extremely sensitive to choice of the \xco\ value,  since the molecular gas is the 
dominant component in the inner parts. 
 We found that for three galaxies of our sample, namely NGC4254, NGC4321, and NGC4501, 
the slopes of the oxygen and of the dust-to-gas radial gradients  agree up to $\sim$0.6-0.7R$_{25}$  using  \xco\  values in the range 1/3-1/2 Galactic \xco.
For NGC4303 a lower value of \xco$\sim$0.1$\times$ 10$^{20}$ is 
necessary. }
{We suggest that such low  \xco\ values might  be due to a metallicity dependence of \xco 
(from close to linear for NGC4254, NGC4321, and NGC4501 to superlinear for NGC4303), 
especially  in the radial regions R$_G<$0.6-0.7R$_{25}$
where the molecular gas dominates. On the other hand, the outer regions, 
where the atomic gas component is dominant, are less affected by the choice of \xco, and thus we cannot put constraints 
on its value.}
{}
\authorrunning{Magrini, L. et al.}
\titlerunning{Dust and metals  gradients in Virgo spirals}

   \maketitle
%

\section{Introduction}
Virgo is one of the best studied galaxy clusters, being the richest cluster nearest to our own Galaxy 
\citep[$\sim$17 Mpc,][]{1999MNRAS.304..595G}. 
It is a relatively populous system, consisting of more than 1000
confirmed members  \citep{1985AJ.....90.1681B}.
Galaxies in clusters such as  Virgo differ
significantly from their field counterparts since interactions 
with the hostile environment remove gas, quenching the star formation  
\citep[c.f.,][]{2006PASP..118..517B}.

A galaxy's metallicity is closely related to the star formation (SF) history by which
the interstellar medium (ISM) is enriched with the end-products of stellar evolution, 
and to the infall process that dilutes the ISM and triggers new SF. 
As a consequence of their star formation histories, gas stripping and infall events, modified by the cluster environment,  
galaxies in clusters are expected also to differ in metal content relatively to isolated galaxies.   
A fundamental tool with which  tracing the chemical evolution of a galaxy is the study of its 
radial metallicity gradient. The metallicity gradient tracks indeed the star formation history of galaxies, 
integrated over time, together with infall and/or outflow events. 

The first pioneering work in Virgo was done by \citet{1996ApJ...462..147S}, who analyzed nine spiral
galaxies with the aim of seeking correlations among their gas content, locations in the cluster,
metallicities and radial gradients, and comparing them with field spirals. 
\citet{1996ApJ...462..147S} found weak evidence of
shallower gradients in cluster galaxies deficient in HI than gradients in galaxies with a normal HI content. The situation is however very complex because galaxy
interactions affect the star formation history and the gas content across the disk, producing 
metallicity gradients which differ from those measured in isolated galaxies 
of the same morphological type. 
 \citet{2010ApJ...723.1255R} have
shown that gradients in strongly interacting galaxies are flatter than in similar isolated galaxies; on the
other hand, the metallicity gradient of M81, the largest member of a small group of galaxies, is steeper
than in an isolated counterpart due to gas removal in the outskirts 
\citep{2010A&A...521A...3S}.
 
A direct correlation between gas metallicity and the dust-to-gas mass ratio is naturally expected 
since approximatively half of the metals in the ISM reside in dust grains; 
thus the dust-to-gas ratio should scale with metal abundance. 
Such a trend has been obtained theoretically by models computing consistently 
the evolution of metals and dust, despite the large uncertainties in the yields of both 
\citep{1998ApJ...501..643D,2003PASJ...55..901I}.
The relation of the global dust-to-gas mass ratio with metallicity was investigated by, e.g., 
\citet{2002MNRAS.335..753J,2007ApJ...663..866D,2008PASJ...60S.477H,1998ApJ...496..145L}.
The radial variation of the dust-to-gas mass ratio was first investigated by 
\citet{1990A&A...236..237I} in our Galaxy and in other nearby galaxies  (LMC, SMC, M31, M33, and M51)
who found  evidence for a correlation, 
with dust-to-gas mass ratio and metallicity decreasing at roughly the same rate with increasing
galactocentric radius.  
More recently, \citet{2004A&A...424..465B,2005ApJ...619L..83B} and 
\citet{2007ApJS..173..572T}  found a clear relationship between metallicity and 
extinction, thus dust, in several nearby galaxies, 
suggesting that the variation in extinction is associated with the metallicity gradient. 
\citet{2009ApJ...701.1965M} found a good correlation between dust-to-gas
and metallicity gradients in the Spitzer Infrared Nearby Galaxies Survey 
\citep[SINGS;][]{2003PASP..115..928K}.
Finally, \citet{2010MNRAS.402.1409B} compared the  dust-to-gas ratio and metallicity gradients in NGC2403, 
finding a similar decreasing behavior with radius.

We have recently obtained observations of the Virgo galaxies with the 
Herschel Space Observatory \citep{pilbratt10},
within the Open Time Key Project HeViCS (Herschel Virgo Cluster Survey)
\citep{2010A&A...518L..48D}.
HeViCS maps a wide area over the Virgo Cluster at wavelengths from 100 to 
500~$\mu$m. This spectral range covers the peak of the thermal emission 
from cold dust (T$<$30 K) which enables the detection of the bulk of 
the dust emission in galaxies. Also, Herschel gives an unprecedented 
resolution at these wavelengths (ranging from about 10 to 36$\arcsec$, 
equivalent to 1-3~kpc for Virgo galaxies). 
Atomic and molecular gas maps are available in the
literature at a comparable resolution 
\citep{2009AJ....138.1741C,2007PASJ...59..117K},
thus providing resolved maps and the
possibility of deriving radial gradients of dust-to-gas mass ratios. 

In the present paper we assessed the validity of using HeViCS 
observations to obtain metallicity gradients from radial profiles of the dust-to-gas mass ratio. 
We investigated  the hypothesis that the local dust-to-gas mass ratio is
proportional to metallicity, starting from the  four  spiral galaxy in the Virgo cluster  
(NGC4254, NGC4303, NGC4321, and NGC4501)
whose metallicity gradients are available in the literature.
We studied the relation between radial profiles of metallicity and 
dust-to-gas mass ratios, and how this can be used to constrain the 
CO-to-H$_2$ conversion factor (\xco) and its dependence on metallicity. 

The paper is structured as follows:
in Sect. 2 we describe the new HeViCS observations, and the observations of atomic
and molecular gas. We also derive the oxygen abundance and its radial gradient in each
galaxy. Sect. 3 discusses the fits to the observed Herschel dust spectral energy distributions (SEDs),
how the radial profiles are derived, and how we calculate dust and gas masses. 
In Sect. 4 we develop the method by which we constrain \xco, and
Sect. 5 gives our conclusions.

\section{Data}

\subsection{The sample}

In the HeViCS field there are four galaxies for which the oxygen gradient has been well 
determined in the literature \citep[e.g.,][]{1996ApJ...462..147S,2010ApJS..190..233M}. 
They are NGC4254 (M99), NGC4303 (M61), NGC4321 (M100), and NGC4501 (M88), 
all Sbc late-type galaxies. 

NGC4254 is a bright spiral galaxy located at the periphery of the Virgo cluster, 
at a projected distance of $\sim$1~Mpc from the cluster center.  
Optical images show that this galaxy has one-armed structure, also seen in the HI gas distribution 
\citep{1993ApJ...418..113P,2009AJ....138.1741C}.
Such an asymmetric spiral pattern is often observed in tidally galaxies, 
but there is no apparently  massive companion near NGC4254 
\citep{2003PASJ...55...75S}.\\
NGC4321 is located at a distance of $\sim$1.1~Mpc form M87, and has an HI disk that is slightly larger 
than the optical disk. \\
NGC4501 is the closest galaxy of our sample to M87, being located at a  distance of $\sim$0.5~Mpc. 
It is  weakly HI-deficient, following the definition of 
\citet{2009AJ....138.1741C}.
Comparisons with simulations suggest that NGC4501 is 
in an early stage of ram pressure stripping 
\citep{vollmer08},
entering the high-density region of the cluster for the first time.\\
NGC4303 is the most isolated galaxy in our sample. It is a barred spiral galaxy with face-on geometry located in the outskirts of the Virgo cluster. 
We assume for all the galaxies  the distance of $\sim$17~Mpc 
\citep{1999MNRAS.304..595G}.

\subsection{Herschel observations}

The HeViCS program consists of Herschel observations of an area
of about 60 sq. deg. over the denser parts of the Virgo Cluster.
The total area is made of 4 overlapping fields, which
are observed in parallel scanning mode ({\em fast} scan rate: 
60$\arcsec$/s) with both the PACS and SPIRE instruments,
yielding data simultaneously in 5 spectral bands, at 100
and  160~$\mu$m (from PACS) and at 250, 350, and 500~$\mu$m (from SPIRE).
At the completion of the program, each field will be covered
with 8 scans done in two perpendicular scan directions.
The full width half maximum (FWHM) of the beams is 6\farcs98 $\times$ 12\farcs7
and 11\farcs64 $\times$ 15\farcs65 in the two PACS bands
(PACS Observers' Manual, 2010), and 18\farcs2, 24\farcs9, 36\farcs3
in the three SPIRE bands (SPIRE Observers' Manual, 2010).
At the time of writing, each field has been observed with at least
two scans. In this paper, we use this dataset, whose data
reduction and analysis is described in details in PAPER VIII, \citet{davies11}.

 Some papers presenting  Herschel observations for  these galaxies  have been already published. 
\cite{2010A&A...518L..62E} and  \cite{2010A&A...518L..64S} analyzed the SPIRE maps of 
NGC4254 (M99) and NGC4321 (M100)
observed within the Herschel Reference Survey 
\citep{2010PASP..122..261B} to  map the ISM using dust emission. Adding these data to archival Spitzer, HI, and CO maps, 
\cite{2010A&A...518L..72P} investigated the spatial distribution of gas and dust in these same galaxies.
They also present as a preliminary result, the ratio of the total gas mass 
(${\rm HI}+{\rm H}_2$) to 500 $\mu $m flux, an approximation of the dust mass for the two galaxies. 
They found a decreasing dust-to-gas mass ratio with radius, 
consistent with results  by, e.g.,  \cite{2010MNRAS.402.1409B} in NGC2403. 
With the  present-time availability of the PACS data  the  dust SED fitting con be 
better defined allowing to  measure the exact shape of the radial 
dust-to-gas mass gradient. 
Finally, \cite{2010A&A...518L..51S} presented a resolved dust analysis of three of the 
largest (in angular size) spiral galaxies in HeViCS, among them NGC4501.

\subsection{The calibration of metallicity and the abundance gradients}
\label{sec_grad}
The metallicity measurements of these galaxies are available in the literature 
from optical spectroscopy of their HII regions.
The most direct method to derive the oxygen abundance is to measure the electron temperature (T$_e$) of the ionized gas using the intensity (relative to a hydrogen recombination line) of one or more temperature-sensitive auroral lines such as [O III] $\lambda$4363\AA, [N II] $\lambda$
 5755\AA, [S III] $\lambda$6312\AA, and [O II] $\lambda$7325\AA, as summarized by 
\cite{2010ApJS..190..233M}.
However, measurements of the electron temperature were not available from the spectroscopic observations in the original papers of \cite{1985ApJS...57....1M}, \cite{1991ApJ...371...82S},  \cite{1994MNRAS.266..421H}, and 
\cite{1996ApJ...462..147S}. 
The temperature diagnostic  lines are indeed intrinsically faint in metal-rich HII regions. 
Therefore, oxygen abundance has been  derived using  the  strong-line abundance calibrations which relate 
the metallicity  to one or more line ratios involving the strongest recombination and forbidden lines. 
In particular,  the oxygen excitation index R$_{23}$=([OII]+[OIII])/H$_{\beta}$ is one of the most often adopted calibrators to estimate the nebular abundances.
As discussed by \cite{2010ApJS..190..233M}, the principal advantage of R$_{23}$ as an 
oxygen abundance diagnostic is that it is directly proportional to both principal ionization 
states of oxygen, whereas one of the major disadvantages is that 
the relation between R$_{23}$ and metallicity is degenerate  for low and high metallicity. 

However, the calibrations of the metallicity by means
of the strong-line ratios are not unique. They  can be divided in  three main categories: those calibrated with photoionization  models
\citep[e.g.,][]{1991ApJ...380..140M,1994ApJ...420...87Z,2002ApJS..142...35K,2004ApJ...617..240K};
those calibrated directly with the electron temperature, called empirical methods 
\cite[e.g.,][]{2001A&A...374..412P,2004MNRAS.348L..59P,2005ApJ...631..231P};
and those combining both methods 
 \citep[e.g.,][]{2002MNRAS.330...69D}.
As discussed widely in \citet{2008ApJ...681.1183K}, 
the abundances derived with different methods do not have a common absolute oxygen abundance scale. 
The oxygen abundances derived using the theoretical calibration are up to  a factor of $\sim$4 higher than those based on the empirical calibration. 
However, despite the significant zero-point offset in the abundance scales, 
to first order the slope of the abundance gradients agree when calculated with different 
calibrators \citep[see][]{2010ApJS..190..233M}.

We have tested the effect of several metallicity calibrators in NGC4254, 
the galaxy with the best sampled metallicity gradient. 
Starting with the abundance estimates from \citet{2010ApJS..190..233M}, 
based on literature spectroscopy  calibrated with the  \cite{2004ApJ...617..240K} formula (hereafter KK04), 
we used the relationships provided by \citet{2008ApJ...681.1183K} to convert to other abundance scales. 
The relationships of \citet{2008ApJ...681.1183K} are obtained  for limited oxygen abundance ranges, 
corresponding to the ranges where they could perform a polynomial fit 
to transform one abundance scale to another.  
Because of this, from 
the abundances of KK04 we were unable to recover 
the oxygen determination of \citet{2004MNRAS.348L..59P} and \citet{2005ApJ...631..231P}. 
For these cases, we recomputed the oxygen abundance from the original spectra 
\citep{1985ApJS...57....1M,1991ApJ...371...82S,1994MNRAS.266..421H}.

\begin{figure}[htbp]
\centering
\includegraphics[width=90mm]{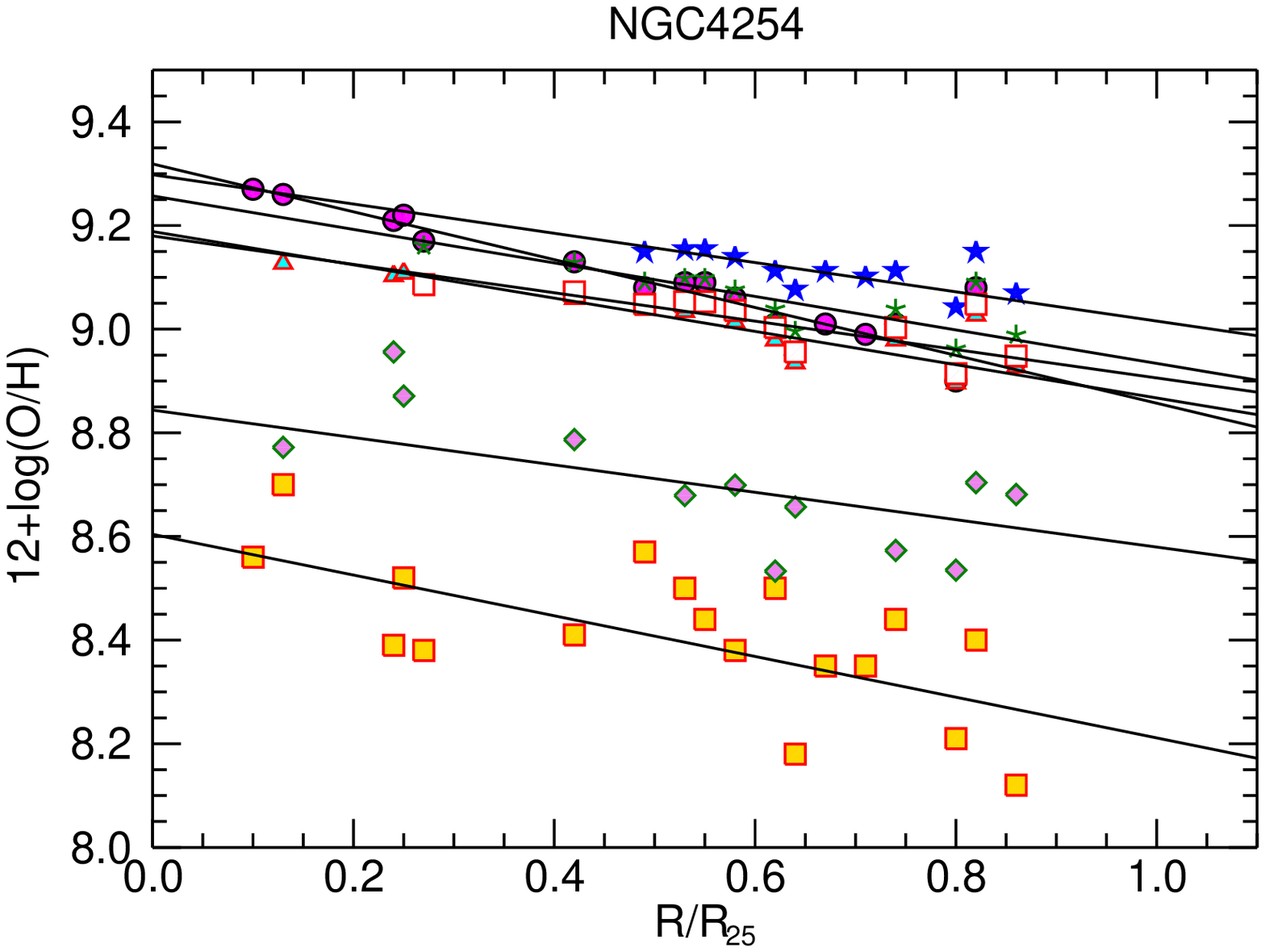}
\caption{The oxygen abundance gradient of NGC4254 obtained with several metallicity calibrations:  \cite{2004ApJ...617..240K} (magenta circles),  \cite{2005ApJ...631..231P} (yellow squares), 
\cite{1994ApJ...420...87Z} (cyan triangles), \cite{1991ApJ...380..140M} (blue stars), \cite{2002ApJS..142...35K} (empty squares), \cite{2004ApJ...613..898T} (green asterisks), 
and \cite{2004MNRAS.348L..59P} (violet diamond).
}
\label{fig_allmet}
\end{figure}

In Fig.~\ref{fig_allmet}, we show the result of our test: while the slope of the 
gradient is almost invariant with different calibrations, the zero-point depends on the choice of the 
calibration.   
In particular, the empirical calibrations of \citet{2004MNRAS.348L..59P} and 
\citet{2005ApJ...631..231P}  show values of the oxygen abundances roughly 0.5-0.8~dex lower
than the other determinations. 
These two latter empirical calibrations were obtained for HII regions with available 
electron temperature in a relatively low metallicity regime. They could not be valid 
for metal rich environments, as the galaxies of our sample. 
In Sect.~\ref{scales} we will discuss how the dust-to-gas ratio might help in setting a 
lower limit to the metallicity and to discriminate among different calibrations. 
  
\subsection{HI and CO maps}

The radial profile of the gas, including both atomic and molecular components, 
is necessary to derive the dust-to-gas mass ratio gradient. 
For NGC4254, NGC4321, and NGC4501, we use the moment-0 HI maps obtained with VLA Imaging survey of Virgo
galaxies in Atomic gas (VIVA) survey by \citet{2009AJ....138.1741C}. 
VIVA observations reach a column density sensitivity 
of 3-5$\times$10$^{19}$cm$^{-2}$.
The comparison of their total HI fluxes with values in the 
literature from single dish observations gives a  good agreement 
especially for the large galaxies, indicating 
no loss of flux in the interferometric observations 
\citep[see Fig.~5 in][]{2009AJ....138.1741C}.
The beam sizes are: 30\farcs86$\times$ 28\farcs07 for NGC4254, 
15\farcs90$\times$ 14\farcs66 for NGC4321, 
and 16\farcs83$\times$ 16\farcs41 for NGC4501.
The  HI radial profile of NGC4303 
is available from \citet{1988A&AS...73..453W} and \citet{1990AJ....100..604C}. 
We adopt the combined radial profile of the two, shown in Fig.3 of 
\citet{1996ApJ...462..147S}. 

Maps of molecular gas were available thanks to the $^{12}$CO ($J$ = 1--0) 
mapping survey of 40 nearby 
spiral galaxies, performed with the Nobeyama 45~m telescope by \citet{2007PASJ...59..117K}.

\section{Analysis}

\subsection{The mass and temperature of dust}
\label{dust}
Maps of the dust temperature and mass surface density were obtained as in 
\citet{2010A&A...518L..51S}. The images of the galaxies from the 
five PACS (100 and 160~$\mu$m) and SPIRE (250, 350, and 500~$\mu$m) bands 
were all convolved and re-gridded to the lower resolution 
(FWHM=36\farcs9, 3~kpc) and pixel size 
(14\farcs0, 1.1~kpc) of the 500~$\mu$m observations.
We used only pixels with S/N$>10$ at 500~$\mu$m 
\citep[the {\em rms} of our maps in this band is about 0.3 MJy/sr, see][]{davies11}.
The selection of these high surface-brightness pixels was necessary 
to limit the uncertainties 
due to background subtraction and avoid the artefacts caused by the 
high-pass filtering in the PACS data reduction. 
Despite this limit, we were able to study the dust and gas properties
up to at least 0.7 R$_{25}$\footnote{R$_{25}$ is the radius of the galaxy measured to a B surface brightness of 25 mag arcsec$^{-2}$, and is an indication of the size of the galaxy; R$_{25}$ were obtained from NED; $\sim$0.7R$_{25}$  is equivalent to the solar radius in our Galaxy. }  in all galaxies.
For each galaxy we thus considered  approximately 150-200 pixels covering about
1/2 of the optically defined area.

We estimated the error on the surface brightness
on a pixel by pixel basis by comparing 
galaxy images from the two scan data used in this  paper  
with those relative to other two scans recently taken by Herschel
which cover only part of the HeViCS field.  
We found errors very similar to those estimated on the total fluxes 
\citep[see][]{davies11}.
Including a calibration error of 15\% for PACS (PACS ICC, private communication) and 7\% for SPIRE (SPIRE Observer' Manual, 2010), the 
total error is 30\%, 20\%, 10\%, 10\%, and 15\% of the flux at 
100, 160, 250, 350, and 500~$\mu$m, respectively.

The SED for each pixel was fitted with a single modified blackbody, using 
a power law dust emissivity $\kappa_\lambda=\kappa_0 (\lambda_0/\lambda)^\beta$,
with spectral index $\beta=2$ and emissivity $\kappa_0$ = 0.192 $m^2$ kg$^{-1}$ 
at $\lambda_0$ = 350~$\mu$m. These values reproduce the average emissivity of 
models of the Milky Way dust in the FIR-submm \citep{2003ARA&A..41..241D}.
The fit was obtained with a standard $\chi^2$ minimization technique. 
In the pipeline calibration, the flux density observed by the various 
instruments, i.e. weighted over each filter passband, is converted
into a monochromatic flux density assuming $F_\nu \propto \nu^{-1}$
Before fitting a modified blackbody, a color correction 
should be applied to the data, to account for the real
spectral slope of the source. Alternatively, the conversion implemented into
the pipeline calibration can be removed from the data 
(in SPIRE parlance this is equivalent to dividing the pipeline flux densities by the $K_4$ 
factor; SPIRE Observer' Manual, 2010); the passband weighted flux thus 
obtained should then be compared with the mean of the model flux density 
over the spectral response function for each of the bands. We adopted
this second technique, using the appropriate response functions
for the PACS and SPIRE bands (for SPIRE, we used the response functions
for extended emission).  However, color corrections are small
for the adopted emissivity and the temperature range derived here
\citep[see][]{davies11}.

\begin{figure}[htbp]
\centering
\includegraphics[width=90mm]{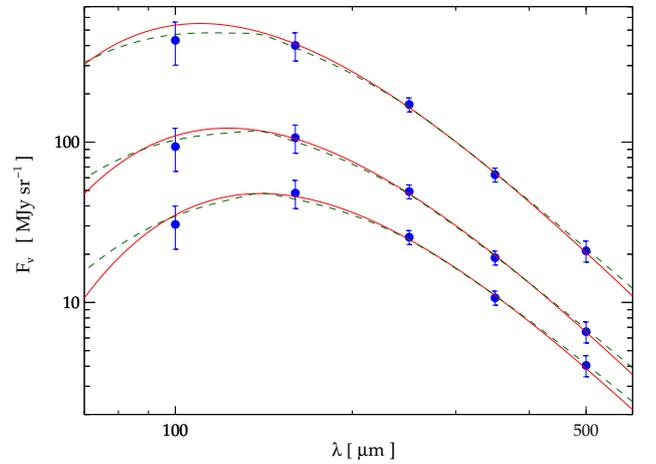}
\caption{Typical SEDs on three different positions (pixels) on NGC4254.
Blue dots are the measured (color-corrected) fluxes and the red (solid) +
curves are the modified blackbody fits.  The three SED correspond to minimum
($\approx 21K$), mean ($\approx 23.5K$), and maximum ($\approx 26K$)
temperature in the galaxy (see Fig.~\ref{dust_t}). The green (dashed) curves
show the fit of the \citet{2007ApJ...657..810D} model to the data, for IRSFs
of intensities 2, 4 and 8 times the Local ISRF (see their paper for details).}
\label{fig_sed}
\end{figure}

In Fig.~\ref{fig_sed} we show typical SEDs and graybody fits for three positions
of different dust temperature in NGC4254.
As shown by the figure, a single temperature modified blackbody with $\beta=2$ is
sufficient to obtain reasonable fits of the SED over the wavelength
range considered here \citep[see also][]{davies11}.
When data at
shorter wavelength than PACS 100~$\mu$m is available, one might want
to consider a two-component model, to include emission from warmer
dust that might significantly contribute at least to the 100~$\mu$m
flux. This was done, for example, by \citet{2010A&A...518L..65B}, who used Herschel data,  and by \citet{2010A&A...518L..51S} who
used 70~$\mu$m data from the Spitzer satellite. However, they found
that the inclusion of a warmer temperature component, thought 
necessary to fit the  70~$\mu$m data, improves the fit at
$\lambda > 100$~$\mu$m only slightly, and does not modify significantly 
the estimate of the temperature and mass surface density of cold dust.

\subsection{Sources of uncertainty in mass and temperature of dust}

When only the errors on photometry are considered, the uncertainty in the 
determination of the temperature is about 2~K, and thus $\sim$20\% on the 
dust mass surface density. 

In principle, fitting the dust SED with a single thermal-equilibrium
temperature component could result in larger uncertainties in the
dust mass estimates: grains of a given size and material could be exposed 
to different intensities of the interstellar radiation field (ISRF) and thus
attain different equilibrium temperatures which will contribute differently
to the SED; conversely, for the same radiation field the SED could depend 
on the dust distribution, because it results from the emission of a mixture 
of grains of different size and composition, each with its own equilibrium 
temperature. We found neither of these to have a strong effects 
on our mass estimates. In fact, by fitting the SED pixel by pixel, we 
already take into account the gradients due to the diffuse ISRF, which is 
more prominent in the radial direction 
than in the vertical directions (i.e.\ along the line of sight, for non
edge-on disks; \citealt{2000A&A...359...65B}). 

However, the temperature radial gradients found here (see 
Sect.~\ref{sec_profiles}) would not produce very large uncertainties,
even when the global SED is fitted with a single temperature model. For
our targets, the differences between the sum of the dust mass in each
pixel, and the dust mass obtained by fitting the sum of the flux
densities in each pixel, is smaller than the fit error. The insensitivity
of the global dust SED fitting on the shallow diffuse ISRF
gradient is clearly shown in the analysis of the the FIR/submm SED of
late type galaxies of \citet{2007ApJ...663..866D}: though the more
complex fitting procedure includes a dust grain model and a
range of ISRF intensities (see also \citealt{2007ApJ...657..810D}), 
the global SED at $\lambda\ge 100\mu$m
is found to be fitted by a dust component that accounts for most of
the dust mass (a part from a few percents), heated by an IRSF
of constant intensity.

We evaluated the effects of a complex dust mixture by fitting the
model of \citet{2007ApJ...657..810D} to our pixel-by-pixel SED.
Following \citet{2007ApJ...663..866D}, we used a single value
of the ISRF for emission at $\lambda\ge 100\mu$m.
An example can be seen in Fig. ~\ref{fig_sed}. The SED from the
dust grain mixtures and our single temperature model fit equally 
well the data.  The dust mass obtained by using the procedure of
\citet{2007ApJ...663..866D} is higher by about 10\%, that is,
within the error we quoted. Thus, the dust mass derived 
from a simple averaged emissivity, and a single IRSF - or
temperature - , is not severely underestimated.

Larger uncertainties can come also from the assumption
for the emissivity. For example, the value of the emissivity
derived by \cite{2002MNRAS.335..753J} from SCUBA observations of galaxies
is equivalent to $\kappa_0$ = 0.41 $m^2$ kg$^{-1}$, a factor of two larger
than the value we adopted. Adopting this emissivity would result in dust
surface densities a factor of two smaller than what we found in this paper.

Also, dust emissivity has been reported to increase by about a factor two
in grains associated to denser environments
\citep[see, e.g.,][]{2003A&A...399L..43B}. However, this might be the
case for extreme environments and not representative of the bulk of the
diffuse dust mass: in the Milky Way,
recent results from the Planck satellite show no emissivity
variation betweed dust associated with HI and CO emission, nor emissivity 
variations with the galactocentric radius \citep{2011arXiv1101.2032P}.
In Sect.~\ref{scales} we will discuss how the uncertainty on $\kappa_0$
might affect our discussion.

The limited wavelength coverage does not allow us to investigate in
details the effect of variation of the emissivity spectral index.
In any case, the modified $\beta=2$ blackbody provides good
fits for all the SEDs analised here, as well as for the
global SEDs in a larger sample of HeViCS objects \citep{davies11}.
Local variations of the dust emissivity index with temperature, as
those reported by \citet{2010A&A...520L...8P} cannot be easily veryfied 
in our dataset. Variation of $\beta$ from 1.7 to 2.2, as those
measured at a reference temperature of 20K, would result in an
underestimate and overestimate of the dust mass of the same order as
the quoted errors, respectively. Lower $\beta$ values, as those
found for dust at higher temperatures, will not be able to provide
good fits to our SEDs in the central part of the galaxies.

Maps of the temperature and dust mass surface density for
NGC4501 have been presented in  \cite{2010A&A...518L..51S}. For the other
galaxies, they will be  presented in Vlahakis et al. (2011, in preparation).
\begin{figure} 
\resizebox{\hsize}{!}{\includegraphics{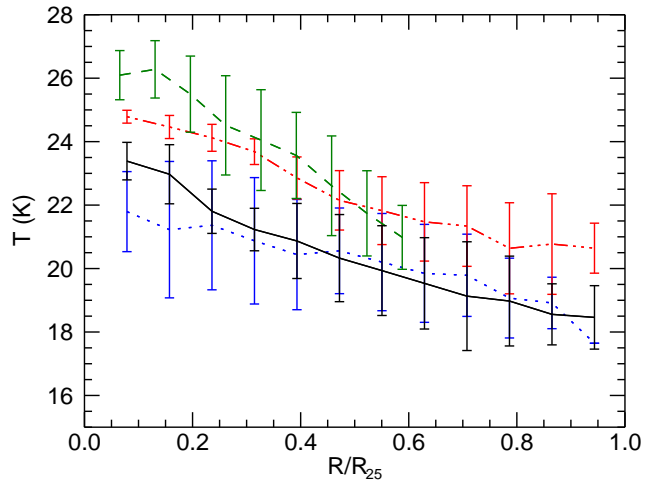}} 
\caption{Radial variation of dust temperature in NGC4254 (red dot-dashed curve), NGC4303 (green dashed line), 
NGC4321 (blue dotted curve), and NGC4501 (black solid curve).  }
\label{dust_t} 
\end{figure}

\subsection{Radial profiles}
\label{sec_profiles}
 
The galaxies in our sample are disturbed by tidal interactions, thus we consider more 
reliable the determination of the geometric parameters of the galactic disks
from kinematics rather than from photometry
\citep[see, for example, the discrepancy between parameters derived with the two methods
for the case of NGC4321 in][]{2009AJ....138.1741C}. 
The task {\sc ROTCUR} of the package {\sc NEMO} 
\citep{1995ASPC...77..398T}
was used to fit
a tilted ring model to the 21-cm moment-1 maps from the VIVA dataset. We obtained
values for the rotational velocity, the inclination and the position angle, as a 
function of the galactocentric radius. 

\begin{figure*}[htbp]
\centering
\includegraphics[width=120mm]{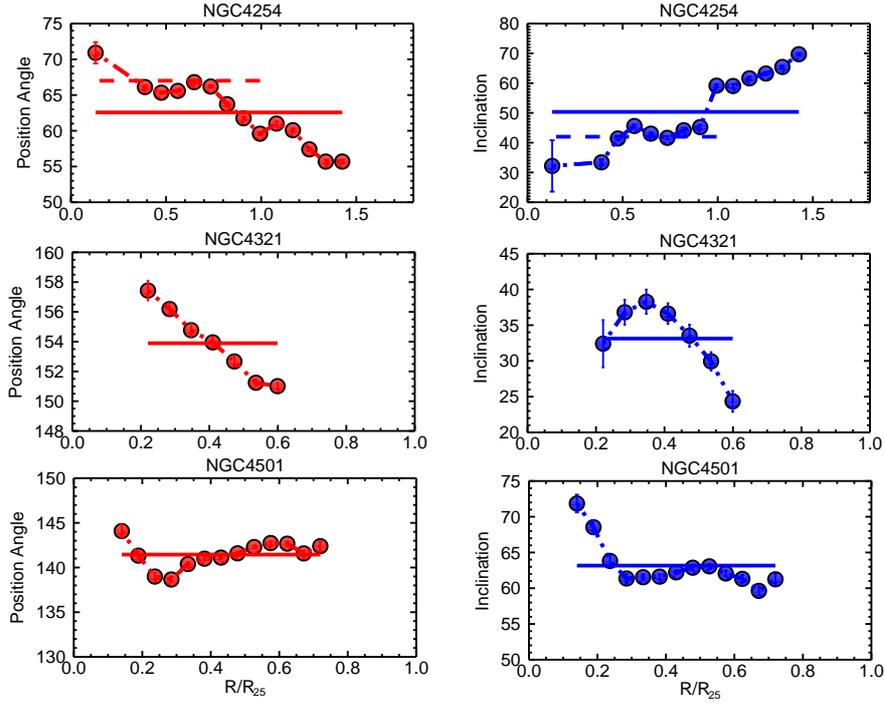}
\caption{Radial variation of position angle and inclination in NGC4254, NGC4321, and NGC4501 from moment-1 H~I maps. The continuous lines are the average in the whole radial range, whereas the 
dashed line  is the average in the region  $R_{G}<R_{opt}$ for NGC4254.}
\label{geo}
\end{figure*}

The radial variations of {\em i} and p.a. are shown in Fig.~\ref{geo}. The average values 
are indicated with solid lines. 
For NGC4321 and NGC4501,  {\em i} and p.a. are quite constant with radius and thus 
we use the average parameters in the whole radial range. 
For NGC4254, which is a one-armed spiral,  
the variations of p.a. and {\em i} are important due to its asymmetric shape.  
Since in our comparison between metallicity and dust-to-gas mass ratio gradients we are interested mostly 
in the inner regions where the metallicity data are available, for NGC4254 we used {\em i} and p.a. averaged in the regions with 
galactocentric radius $R_{G}<R_{25}$ (dashed line in Fig.~\ref{geo}, top panels) 
where the variations are small. 
\citet{2010A&A...518L..72P} also found different ellipse parameters  for the outer parts and for the inner parts of NGC4254. 
For NGC4303, for which VIVA maps are not available,  we adopted the parameters by \citet{2006PASJ...58..299K}. 

The final values are shown in  Table~\ref{tab:sample}: in column 1 we show the galaxy name, in columns 2 and 3 
the mean {\em i} and p.a., in column 4 the optical diameter $D_{opt}$ in arcmin, in cols. 6, 7, and 8 
the slope of the metallicity gradient, the metallicity at the equivalent solar radius 0.7R$_{25}$, and their reference. 
For NGC4254 we report the value we adopt, i.e. 
the average within the optical radius, and in square brackets the average in the whole radial range. 
Between square brackets, for all galaxies, we report also inclinations and position angles from the literature, 
which are generally in good agreement with our values in the radial range considered.

\begin{figure*}[htbp]
\centering
\includegraphics[width=200mm]{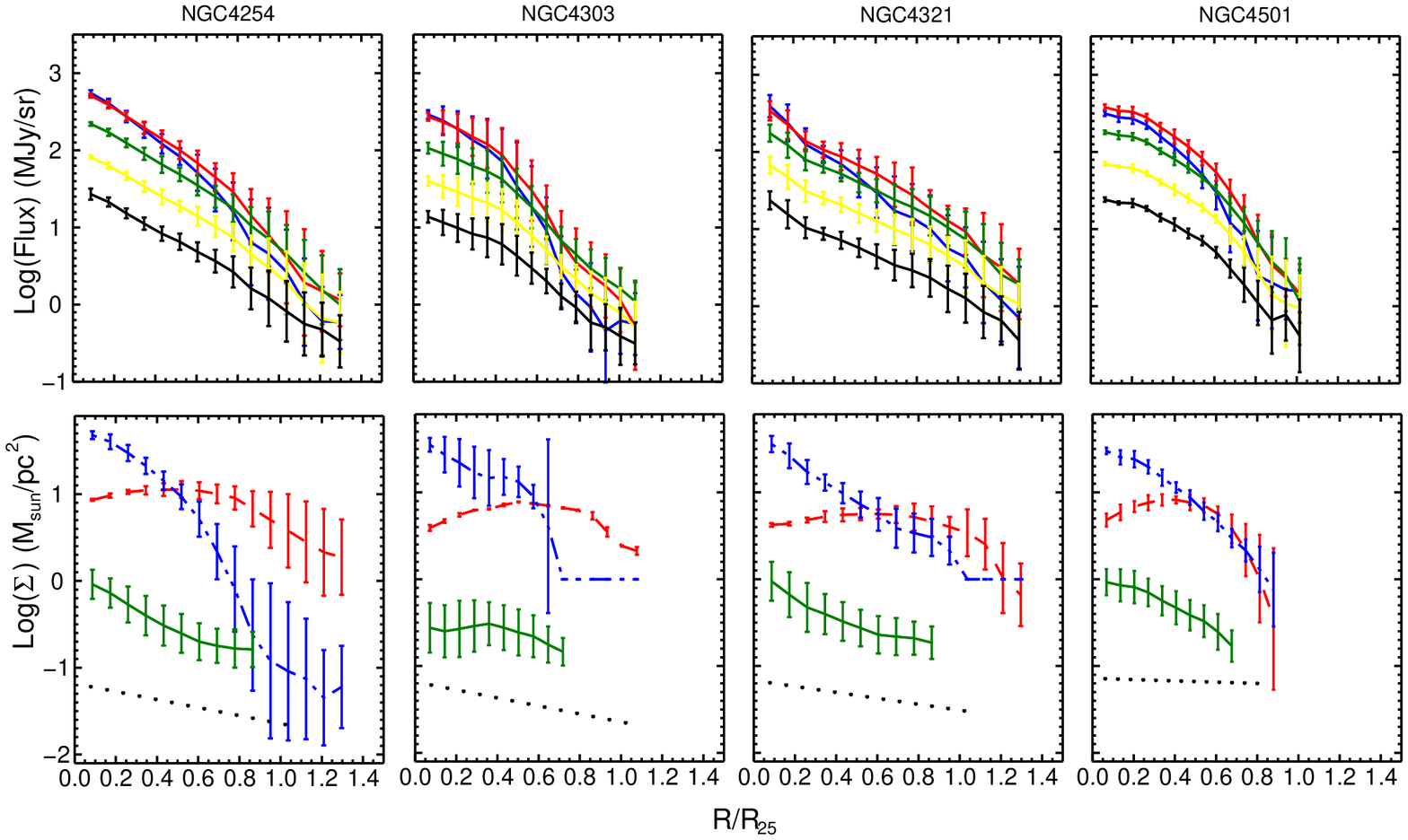}
\caption{\textit{Top panels}: radial profiles of  dust emission at 100, 160, 250, 350, 500~$\mu$m 
(from top to bottom) for NGC4254, NGC4303, NGC4321, NGC4501. \textit{Bottom panels}: radial 
profiles of dust mass(green solid line), HI (red long dashed line), H$_{2}$ (blue dot-dashed line) surface density using a uniform conversion factor 
\xco$\approx (1.8\pm 0.3) \times 10^{20}$ cm$^{-2}$, and the O/H radial gradient (dotted line, scaled of a factor 10.5 to match the logarithmic scale of the other quantities).  
\label{profile_all}}
\end{figure*}

\begin{table*}
\caption{Parameters of the sample galaxies.}
\label{tab:sample}
\begin{tabular}{lllllllll}
\hline\hline
Name & Inclination & Position angle & D$_{opt}$$^{a}$           &slope$_{KK04}$ & 12+$\log$(O/H)$_{KK04}$ &                     Ref. \\
	 &			&			&(arcmin)		   	& (dex/R$_{25}$) &	at R$_{G}$=0.7R$_{25}$	    	     & (7)\\
	 (1)& (2)& (3) & (4) & (5) & (6) \\
\hline
NGC4254 & 42$^{\circ}$ (53$^{\circ}$)$^{b}$ [42$^{\circ}$]$^{d}$ & 67$^{\circ}$ (61.6$^{\circ}$)$^{b}$ [68$^{\circ}$]$^{d}$       &  5.6        &-0.42$\pm$0.06	 & 8.84$\pm$0.02 &M10$^{c}$\\
NGC4303 & 30$^{\circ}$$^{e}$ & 135$^{\circ}$$^{e}$    & 6.5   &-0.42$\pm$0.02$^{\ddag}$      & 9.24$\pm$0.02  &S96$^{c}$\\
NGC4321 & 33$^{\circ}$ [26$^{\circ}$]$^{f}$ & 154.4$^{\circ}$[155$^{\circ}$]$^{f}$ & 7.6         &-0.35$\pm$0.13	 & 9.04$\pm$0.04  &M10$^{c}$\\
NGC4501 & 63$^{\circ}$[58$^{\circ}$]$^{g}$ & 141.4$^{\circ}$[140$^{\circ}$]$^{g}$ & 7.1         & -0.07$\pm$0.16$^{\ddag}$	 & 9.30$\pm$0.21  & S96$^{c}$ \\
\hline
\end{tabular}
\tablefoot{ 
$a$ D$_{opt}$ is 2$\times$R$_{25}$; 
$b$ Average values for R$_{G}<$R$_{25}$ and, between brackets, throughout the whole radial range. 
$c$ M10=\citet{2010ApJS..190..233M} and S96=\citet{1996ApJ...462..147S} ;
$d$ \citet{1993ApJ...418..113P};
$e$ \citet{2006PASJ...58..299K};
$f$ \citet{1997ApJ...479..723C};
$g$ \citet{1988ApJS...66..261K}.
$\ddag$ We recomputed the gradient slope of NGC4303 and NGC4501 using  data from 
\citet{1996ApJ...462..147S}, converting them with \citet{2008ApJ...681.1183K} 
formula to the KK04 scale, 
obtaining  different values with respect to those quoted in their paper. \\
}
\end{table*}
The atomic and molecular gas maps
were convolved to the same resolution of the SPIRE 500~$\mu$m map and regridded on the
same pixel scale. The radial bin size in the profiles correspond to 1 pixel in the maps
(14$\arcsec$ corresponding to $\sim$1.1~kpc). Adopting the geometric parameters in 
Table~\ref{tab:sample}, we used a dedicated {\sc IDL} 
procedure to obtain radial profiles (azimuthally averaged over elliptical annuli) for
HI, CO, and dust mass surface density. 
The profiles are shown in Figs.~\ref{profile_all}: the top panels contain PACS and SPIRE
profiles at 100, 160, 250, 350, and 500~$\mu$m (from top to bottom) for NGC4254, NGC4303, 
NGC4321, and NGC4501; the bottom panels the profiles of dust (green), HI (red),
and H$_{2}$ (blue) surface density, where the CO flux is converted into H$_2$ using the standard
conversion factor  \xco$\approx ( 1.8\pm 0.3) \times 10^{20}$ cm$^{-2}$ 
(K km s$^{-1})^{-1}$ (see below). 
For all galaxies H$_{2}$ is dominant in the inner-most  regions, while  HI becomes dominant over H$_2$ at  R$>$0.5-0.6 R$_{25}$. 
HI radial profiles are quite flat within R$_{25}$, whereas both H$_2$ and dust profiles are decreasing with radius.  
 
The errors on the radial profiles were estimated by  
a combination in quadrature of the uncertainty in the overall absolute calibration and
of the standard deviation of the fluxes in each radial bin. Typical errors are of the 
order of 10-30\%. 
However, uncertainties for the dust mass are mainly due to the choice of the model parameters and 
can vary by a factor two.

In the next Section, we report the total mass of H$_2$ obtained with the different assumptions on  
\xco\ (Tab.\ref{tab:h2}). 

\subsection{The \xco\ factor}

To  convert $^{12}$CO observation to measure the amount of molecular hydrogen we  
assume a CO-to-H$_2$ conversion factor 
$$X_{\mathrm{CO}\,}=N( \mathrm{H}\,_{2}) / \int I( \mathrm{CO}\,) \, dv$$ 
\citep{1983ApJ...274..231L}.
In our Galaxy \xco $\approx ( 1.8\pm 0.3) \times 10^{20}$ cm$^{-2}$ (K km s$^{-1})^{-1}$ with 
excursions of up to a factor of $\sim$2 over this value, particularly at high latitudes 
\citep{2001ApJ...547..792D}.
\cite{2011MNRAS.tmp...11S} and \citet{leroy09} showed that \xco\ can vary even more, especially 
in low metallicity environments  where it can reach \xco$\sim 100 X_{Gal}$. 

As derived by analysis of  observations by \citet{2008ApJ...686..948B} and by theoretical models 
\citep{2011MNRAS.tmp..110G},
this factor can vary due to effects of abundance, excitation, optical
depth, and cloud structure averaged over a large area. 
In particular, the abundance of the heavy elements has an important impact on the value of 
\xco\ 
\citep{1986ApJ...303..186I,1988ApJ...325..389M}.
Metallicity affects cloud structure  both directly,   
as smaller abundances of C and O translate into lower abundance of CO, 
and indirectly, as a lower  dust-to-gas mass ratio diminishes the H$_2$ 
formation rates and the shielding of molecular gas from the photo-dissociation by ultraviolet radiation.
As summarized by \cite{2008ApJ...686..948B}, there are several observational calibrations of 
\xco\ with the metallicity, $Z$, in the literature, showing a range of behaviors; 
most of the calibrations find an increasing \xco\ with decreasing $Z$ 
\citep[e.g.,][]{1995ApJ...448L..97W,2000MNRAS.317..649B,2003A&A...397...87I,2004A&A...422L..47S}.

The dependence varies from \xco $\sim Z^{-2.5}$  
\citep{2000mhs..conf..293I}
to \xco $\sim Z^{-1}$  
\citep{1995ApJ...448L..97W,1996PASJ...48..275A,2002A&A...384...33B}, while
\cite{2008ApJ...686..948B} do not find any  measurable trend in the range from $8<12+\log$(O/H)$<8.8$  
on the scales of the individual CO-bright giant molecular clouds,  but they did not 
argue for a Galactic \xco\ on large scales. 
\cite{1996A&A...308L..21S} also found a constant \xco\  with metallicity. 
However, few works were dedicated to the metallicity dependence of \xco\  in super-solar regions, 
as for example the study of \citet{1996PASJ...48..275A} 
who investigated \xco\ also in two metal-rich galaxies, namely M31 and M51. The spiral galaxies in Virgo  
are thus of particular importance to verify the assumed dependence on  metallicity  
of \xco\ in  metal-rich environments.  
To evaluate the conversion factor \xco\ at  different metallicities 
and relate it to the dust-to-gas ratio, we compared three different assumptions: 

\begin{itemize}
\item[{\em i)}] {\em Uniform}: we used the standard Galactic value from \citet{2008ApJ...686..948B} 
\xco $\approx 1.8 \times 10^{20}$ cm$^{-2}$ (K km s$^{-1})^{-1}$ 
(hereafter Galactic \xco) throughout the radial range. 
\item[{\em ii)}] {\em Linear metallicity dependence}: we used a linear  metallicity dependence as found 
by 
\citet{1996PASJ...48..275A,2002A&A...384...33B}. 
This is similar to the $-0.67$ slope derived by \citet{1995ApJ...448L..97W}.
We adopted the analytic calibration given by \citet{1995ApJ...448L..97W}, but rely on the 
most recent oxygen solar abundance of \citet{2009ARA&A..47..481A} instead of \citet{1989AIPC..183....1G}.  
Assuming that at the new solar oxygen abundance, 12$+\log$(O/H)=8.69,  
\xco $\approx 1.8 \times 10^{20}$ cm$^{-2}$ (K km s$^{-1})^{-1}$, the  
\citet{2002A&A...384...33B}'s relationship becomes
\begin{equation}
\log(X_{\mathrm{CO}})=-1.0\times(12+\log(\mathrm{O/H}))+8.94
\label{w95}
 \end{equation} 
where \xco\ is expressed in unit of 10$^{20}$ cm$^{-2}$ (K km s$^{-1})^{-1}$. 
\item[{\em iii)}] {\em Super-linear metallicity dependence}: we considered a superlinear dependence on metallicity as 
found by \citet{2000mhs..conf..293I}. 
Re-scaling to the \cite{2009ARA&A..47..481A} solar oxygen abundance, the \citet{2000mhs..conf..293I} 
relationship becomes
\begin{equation}
\log(X_{\mathrm{CO}\,})=-2.5\times(12+\log(\mathrm{O/H}))+21.72
\end{equation}
where \xco\ is again expressed in unit of 10$^{20}$.
\end{itemize}

\section{The method}

 We started from the hypothesis that dust-to-gas mass ratio and O/H are tracing 
the same quantity, i.e. the metal abundance relative to hydrogen. Thus we expected 
that the slope of the two radial gradients agree. 
Within this hypothesis, we attempted to constrain the value of \xco, and eventually its dependence 
on metallicity, using the information contained in the slope of the metallicity gradient. 
In fact, due to the large discrepancy on the zero-point of different calibrations as discussed by, 
e.g., \cite{2008ApJ...681.1183K}, it is difficult 
to place these galaxies on an absolute metallicity scale using only inferences from HII optical spectroscopy. 

Here we summarize the steps, while the details are given in the following sections.  
\begin{itemize}
\item[I] We converted the dust-to-gas mass ratio in oxygen abundances with the  relationship of 
Draine et al.~(2007) (see Eq.~\ref{dtg}) 
\citep[][see Eq.~\ref{dtg}]{2007ApJ...663..866D},
thus obtaining an approximate zero-point for the metallicity scale.  
\item[II]  We computed the  radial profile of the dust-to-gas mass ratio with the Galactic 
\xco.
\item[III] Once fixed the approximate abundance scale, we compared the radial profiles 
of the dust-to-gas mass ratio with the super-linear and linear metallicity dependences of \xco.
\end{itemize}

\subsection{I. The dust-to-gas ratio and metallicity scales}
\label{scales}

If we consider that the abundances of all heavy elements are proportional to the 
oxygen abundance and that all heavy elements condensed to form dust in the same way as in the MW, then the dust-to-gas mass ratio scales proportionally to the 
oxygen abundance 
 (Draine et al.~2007)
\citep{2007ApJ...663..866D}
\begin{equation} 
\frac{M_{\mathrm{dust}}}{M_{\mathrm{H}}} \approx 0.007\frac{(\mathrm{O/H})}{(\mathrm{O/H})_{\odot}}, 
\label{dtg}
\end{equation}
where 0.007 is the dust-to-hydrogen ratio of the MW at solar radius, $\sim$0.7R$_{25}$  with optical radius of our Galaxy of 12~kpc \citep{2000Ap.....43..145R}. 
In eq.\ref{dtg},  we use the observed dust-to-gas ratio 0.0073 estimated from observed depletions in the solar neighborhood instead 
of the value from  dust models, 0.010,  about 40\% larger \citep{2007ApJ...663..866D}, and the solar oxygen abundance  12+log(O/H)$_{\odot}$=8.69 from \citet{2009ARA&A..47..481A}.

A similar relationship was obtained also by  \citet{2002A&A...384...33B}  using the data available for the MW
\citep{1994ApJ...428..638S}, the LMC 
\citep{1982A&A...107..247K}, and the SMC 
\citep{1985A&A...149..330B}.

We converted the dust-to-gas ratio, 
obtained with the Galactic  \xco\ at R$_{G}$=0.7\,R$_{25}$, into 12$+\log$(O/H) with Eq.~\ref{dtg}.  

Before comparing the metallicity derived from dust-to-gas ratio with that derived from nebular oxygen abundances, 
we estimated the effect of  the assumed values of the
mass emissivity coefficient and of the \xco\  on the metallicity derived from Eq.~\ref{dtg}. 
Because the determinations of the total gas mass take place in regions where H$_2$ is not negligible, 
the dust-to-gas ratio depends indeed critically on what is assumed about \xco. 
First, we  checked the value of the dust-to-gas ratio we would  obtain at R$_{G}$=0.7\,R$_{25}$
adopting different \xco: using \xco$\approx 4\times 10^{20}$ cm$^{-2}$ (K km s$^{-1})^{-1}$ we would have 
a dust-to-gas ratio $\sim$0.2 dex lower and using \xco$\approx 0.5\times 10^{20}$ cm$^{-2}$ (K km s$^{-1})^{-1}$ we would 
obtain a dust-to-gas ratio $\sim$0.2 dex higher. Thus an uncertainty of $\pm$0.2 dex is associated to the choice of \xco. 

Then, we verified how the emission coefficient  could affect our discussion. 
In the most unfavorable  conditions we could overestimate (or underestimate) the dust mass by a factor two, 
corresponding to a variation of  O/H derived from  Eq.\ref{dtg} of $\pm$0.3 dex.

In addition, we have that a typical error on the dust-to-gas mass ratio of $\sim$30-50\% (including the flux calibration) translates 
into errors of $\sim$0.10-0.15~dex in O/H. 
Considering that these sources of uncertainty are independent we combine them in quadrature, obtaining a total 
uncertainty of $\sim$0.35~dex.

We compared with 12$+\log$(O/H)  obtained at the same galactocentric distance with 
three calibrations: KK04,  \citet{2004MNRAS.348L..59P} (PP04), and \citet{2005ApJ...631..231P} (P05). 
The results for NGC4254, NGC4303, NGC4321, and NGC4501 are shown in Table~\ref{tab:oh}. 

 \begin{table}
\caption{Oxygen abundance from dust-to-gas mass ratio at 0.7R$_{25}$.}
\begin{tabular}{llllll}
\hline\hline
Name 	 & O/H$_{dust-to-gas}$  & O/H$_{KK04}$  &  O/H$_{PP04}$  & O/H$_{P05}$  \\
           	& 0.7R$_{25}$                   &   &       &              \\
    (1)       	& (2)                     & (3)     				& (4) 			& (5)                   \\
\hline
NGC4254                	&  8.84  &  8.95 & 8.65 & 8.35 \\
NGC4303		         &  9.24   & 9.24	 & 8.94 & 8.64 \\
NGC4321			&  9.04   & 9.10  & 8.80 & 8.40 \\
NGC4501                   &  9.04   & 9.30  & 8.90   & 8.70  \\
\hline
\hline
\label{tab:oh}
\end{tabular}
\end{table}

From the comparison of the metallicity obtained for dust-to-gas ratio with several metallicity calibrations
we found that the calibrations of  \citet{2005ApJ...631..231P} show the largest discrepancy with the metallicity derived from dust 
also if we consider all possible  sources of errors, i.e., uncertainty on the emission coefficient, \xco, flux calibrations, and tend to give lower O/H values.
The calibrations of KK04 and \citet{2004MNRAS.348L..59P} are, 
on the other hand, both consistent within the errors ($\sim$0.35 dex  O/H$_{dust-to-gas}$  and $\sim$0.1 dex for O/H from nebular abundances)
with our findings from the dust-to-gas ratio. 
We conclude that, at the equivalent solar radius (0.7~R$_{25}$), the four galaxies in analysis 
are more metal rich than the MW at the same radius (the solar oxygen abundance is 8.69), thus a perfect environment to test if any dependence 
of \xco\ with the metallicity is in place at high metallicity.

Our result is only in apparent disagreement with  \citet{2007ApJ...663..866D}. Their relationship  
between dust-to-gas and metallicity  is derived theoretically assuming that  the interstellar abundances of all heavy elements 
were proportional to O/H, and their fraction scales as in the MW, and then it is  compared  with the oxygen abundances obtained with the calibration of \citet{2005ApJ...631..231P} 
and the dust-to-gas mass ratio adopting \xco$\approx 4\times 10^{20}$ cm$^{-2}$ (K km s$^{-1})^{-1}$. They found a good agreement with the theoretical relationship within a factor of 2. 
In our analysis we found a better agreement with KK04 oxygen abundances since we are using a different  \xco\, factor and a different value of dust-to-gas in the MW at solar radius: 
the P05 oxygen abundances are indeed on average 2-4 times lower than the KK04's ones, but this is compensated by a \xco\, factor 
two times higher, and the assumed dust-to-gas in the MW for the dust models, $\sim$1.5 higher than the one we used, which comes from observations.

 With our choice of a Galactic \xco$\approx 1.8\times 10^{20}$ cm$^{-2}$ (K km s$^{-1})^{-1}$, a dust-to-gas ratio in the solar neighborhood of 0.0073, 
and the relation in eq.\ref{dtg}, we selected the metallicity scale of  KK04, which best agrees with
the scale derived from the dust-to-gas ratio, and convert the abundances of our galaxies to a common scale. 
For NGC4254 and NGC4321 we used the abundance determined by \cite{2010ApJS..190..233M} 
\citep[based on literature spectroscopy calibrated with the formula of ][]{2004ApJ...617..240K}. 
For NGC4303  and NGC4501 we used the original spectroscopy to calculate oxygen abundance on
the abundance scale of KK04. 

With the calibration of KK04, the abundances at 
0.7R$_{25}$ are all super-solar ranging from $\sim$8.95 to $\sim$9.3, 
as shown in Table~\ref{tab:oh}.
The central oxygen abundances would range from 9.3 to 9.5, 
but we have to consider that they are an extrapolation of the metallicity gradient 
up to the galaxy centre and not a real measurement. In addition, the galaxy disk 
does not extend up to the central regions where the bulge is present.  
These central abundances are comparable with recent results obtained from direct electron 
temperature measurement in the HII regions of M81, where  the extrapolation of the 
metallicity gradient up to the galactic centre gives an oxygen abundance of 9.37 
\citep{2010A&A...521A...3S}, and 
in our Galaxy where the central O/H from the gradient of 
\citet{2006ApJS..162..346R} is 9.2 (determination of O/H with optical spectroscopy).

\subsection{II. The dust-to-gas mass gradient with a standard \xco}
\label{sec_cons}
\begin{figure*}[htbp]
\centering
\includegraphics[width=1.\textwidth]{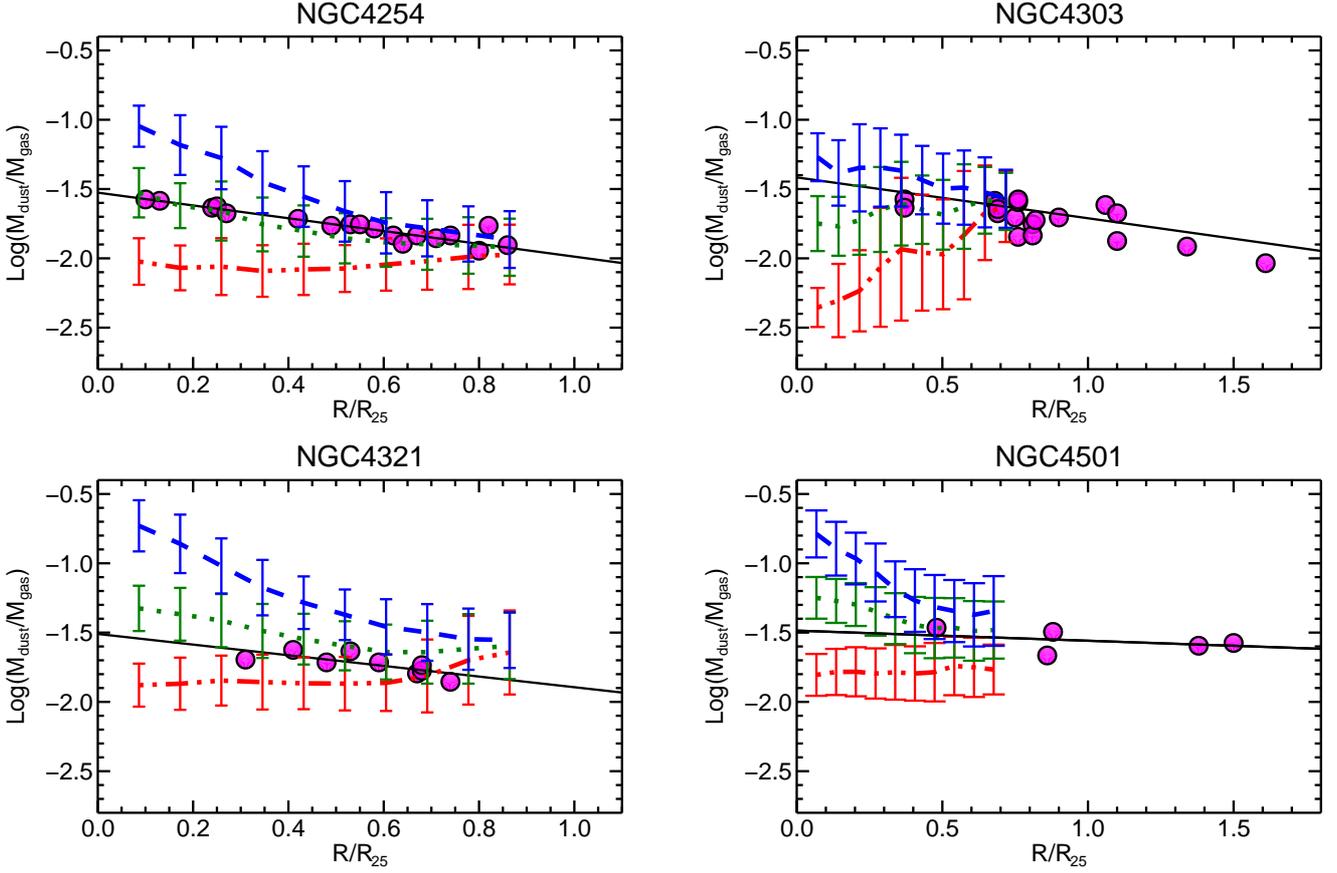}
\caption{Dust-to-gas mass ratio and metallicity radial profiles of NGC4254, NGC4303, NGC4321, and NGC4501 with the 
oxygen abundance calibrated with KK04.  
The conversion factors \xco\ are:  with the Galactic conversion factor  (dot-dashed red curve),  
with the linear  metallicity dependence as in Eq.1 (dotted green curve),  with the superlinear dependence on metallicity as in Eq.~2 (dashed blue curve).  
The magenta circles are the oxygen abundance converted to the dust-to-gas scale with Eq.3.  
For NGC4254 and NGC4321 we used the recalculation by \cite{2010ApJS..190..233M}, for NGC4303 and 
NGC4501 the oxygen abundances by \citet{1996ApJ...462..147S} converted to the KK04 
scale.
The continuous black lines show the fit to the abundance data.  
 \label{f_dtg}}
\end{figure*}

In Fig.~\ref{f_dtg} we show the comparison between the dust-to-gas mass ratio gradients, 
calculated with the Galactic conversion factor (red curves), and the oxygen abundance 
gradients (magenta circles and solid black lines) calibrated with KK04
and  converted in the dust-to-gas mass scale with the relationship discussed in Sect.~\ref{scales}.

If we consider a constant Galactic \xco,  the slopes of the dust-to-gas and metallicity gradients are 
only in marginal agreement. 
In particular, the dust-to-gas gradients are flatter, with a slight positive slope at large 
galactocentric radii R$_{G}>0.7$R$_{25}$.
For NGC4254 and NGC4321 the comparison between the metallicity and dust-to-gas mass ratio 
gradients is straightforward because their metallicity gradients are well determined in the same radial regions 
where the dust-to-gas map is obtained (Fig.~\ref{f_dtg}, left panels). 
For NGC4254 and NGC4321 the dust-to-gas mass ratio gradient are  flatter than the metallicity radial distribution, and they are  constant within the errors, 
with a slightly increasing behaviour in the outer regions. 
The same is true also for NGC4501, while for NGC4303 dust-to-gas mass ratio gradient is positive. 
For these two galaxies we are planning new spectroscopic observations of HII regions in the inner regions  to have a better overlap 
between dust-to-gas and metallicity.  

\subsection{III. Can we constrain \xco? }

We used Eqs. 1 and 2  along with the abundance gradients given in Table~\ref{tab:sample} 
to calculate the molecular gas surface densities, and then the dust-to-gas mass ratios.

The constant Galactic value of \xco\ (dot--dashed red curves in Fig.~\ref{f_dtg}) 
produces in 
most cases flat dust-to-gas mass ratio as discussed in Sect.~\ref{sec_cons}.
The super-linear dependence (dashed blue curves) results in a dust-to-gas radial gradient 
much steeper than the oxygen gradient, especially in the inner regions. 
A linear fit of  the gradient in the logarithm of the dust-to-gas ratio 
[d log(dust/gas)/dR$_{G}$] (Table~\ref{tab:slopes}) suggests that, within $\sim$0.7~R$_{25}$, 
the linear  metallicity dependence is  able to reproduce, within the errors, 
the same radial slope of the metallicity gradient for  NGC4254 and NGC4321. 
A flattening of the dust-to-gas mass ratio gradient is appreciable in the outer regions, 
where the HI component is dominant. 
For NGC4303 a super-linear metallicity dependence of \xco\ is instead necessary to match within 
the errors the two gradients. For NGC4501 
an  \xco\ value  between the Galactic one and the one derived from the linear metallicity dependence is required.

From Fig.\ref{profile_all} we can figure out what is happening:
for a fixed Galactic \xco\ the dust-to-gas ratio would be
constant, as in shown by dot--dashed red curves in Fig.\ref{f_dtg}. 
Once a metallicity dependence of the type described in Eqs. 1 and 2 is introduced in \xco, 
that  depresses the H$_2$ in the centers (which have super-solar metallicity) and
makes the total gas to look more  like the atomic component, which generally has a much more
constant distribution with R$_G$ than the dust. 
Hence, when adopting the metallicity dependences of \xco, we favor the outwardly decreasing gradients in dust-to-gas ratio.

\begin{table}
\caption{Slopes of the O/H and dust-to-gas gradients within 0.7R$_{25}$.}
\begin{tabular}{lllrrr}
\hline\hline
Name 	& slope$_{O/H}$  & slope$_{Gal}$  & slope$_{linear}$  &  slope$_{superlinear}$  \\
   (1)       	& (2)                        & (3)     		        & (4) 			& (5)                \\
\hline
NGC4254&-0.42$\pm$0.06 &  0.07$\pm$0.06  &  -0.55$\pm$0.07 & -1.17$\pm$0.06 \\
NGC4303&-0.42$\pm$0.02 &  1.11$\pm$0.15  &  0.27$\pm$0.13 & -0.44$\pm$0.11 \\
NGC4321&-0.35$\pm$0.13 &  0.09$\pm$0.09  &  -0.52$\pm$0.08 & -1.13$\pm$0.09 \\
NGC4501&-0.07$\pm$0.16 &  0.07$\pm$0.07  &  -0.45$\pm$0.07 & -1.18$\pm$0.12 \\
\hline
\hline
\label{tab:slopes}
\end{tabular}
\end{table}

\begin{figure} 
\resizebox{\hsize}{!}{\includegraphics{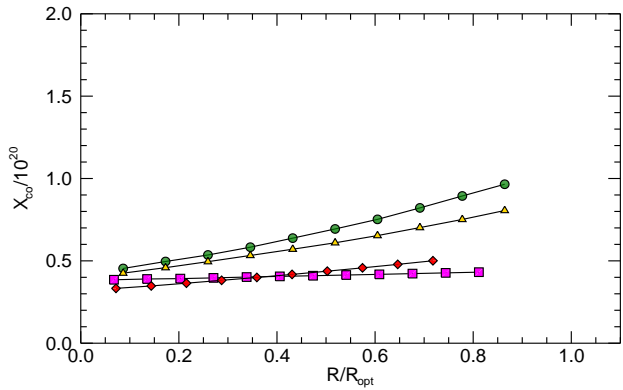}} 
\caption{Radial  variation of \xco\  in the four galaxies with the metallicity computed with the calibration KK04 and the dependence in Eq.~\ref{w95}:  
NGC4254 (green circles), NGC4303 (red diamonds), NGC4321 (yellow triangles), and NGC4501 (magenta squares). }
\label{xco} 
\end{figure}

 Given that these galaxies are dominated by the molecular gas component in their central
regions, one could think that the result we found is, in a certain sense, expected:  we use 
a  \xco\ that depends on 1/Z and we obtain a dust-to-gas mass ratio decreasing as Z. 
However, the radial dependence of \xco\ has only a secondary impact on our results: 
to obtain a decreasing dust-to-gas mass ratio it is necessary only to have a conversion 
factor of the order of $\sim$1/3-1/2 the Galactic value in the central regions of the galaxies, while the outer regions, 
where the atomic gas components is dominant, are less affected by the choice of \xco.
From  Fig.~\ref{xco} where we show the radial dependence of \xco\ 
obtained with Eq.~1 (linear dependence) for NGC4254 (green circles), 
NGC4321 (yellow triangles),  NGC4303 (red diamonds), and NGC4501 (magenta squares), 
it can be seen that no larger radial variations of \xco\  are required. 
To reproduce a decreasing dust-to-gas mass ratio gradient in the radial regions where 
the molecular gas dominates the gaseous component, 
we need for the sample galaxies an \xco$\sim$1/3 of the Galactic value,  
slightly  increasing  with radius  up to  $\sim$2/3  the Galactic value.  
These results agree with \xco\ obtained in the metal rich HII regions 
of M31 and M51 \citep[e.g.,][]{1996PASJ...48..275A}.

To check this, we have used a constant  \xco: 0.5 $\times10^{20}$ cm$^{-2}$ 
(K km s$^{-1})^{-1}$ for  NGC4254, 0.1$\times10^{20}$ for NGC4303,  0.9$\times10^{20}$ for NGC4321, and 0.7$\times10^{20}$ for NGC4501. 
These values are chosen to  produce the best agreement  between O/H and dust-to-gas gradients for each galaxy of our sample. 
They are not related to the \xco\ values derived from the metallicity dependence in Eq.1 and shown in Fig.\ref{xco}. 
Note that the case of NGC4303 seems to be quite extreme: the \xco\ necessary to reproduce with the dust-to-gas ratio the behaviour of the O/H gradient  is much smaller than 1/3 Galactic \xco, being 0.1$\times10^{20}$. 

\begin{figure*}[htbp]
\centering
\includegraphics[width=1\textwidth]{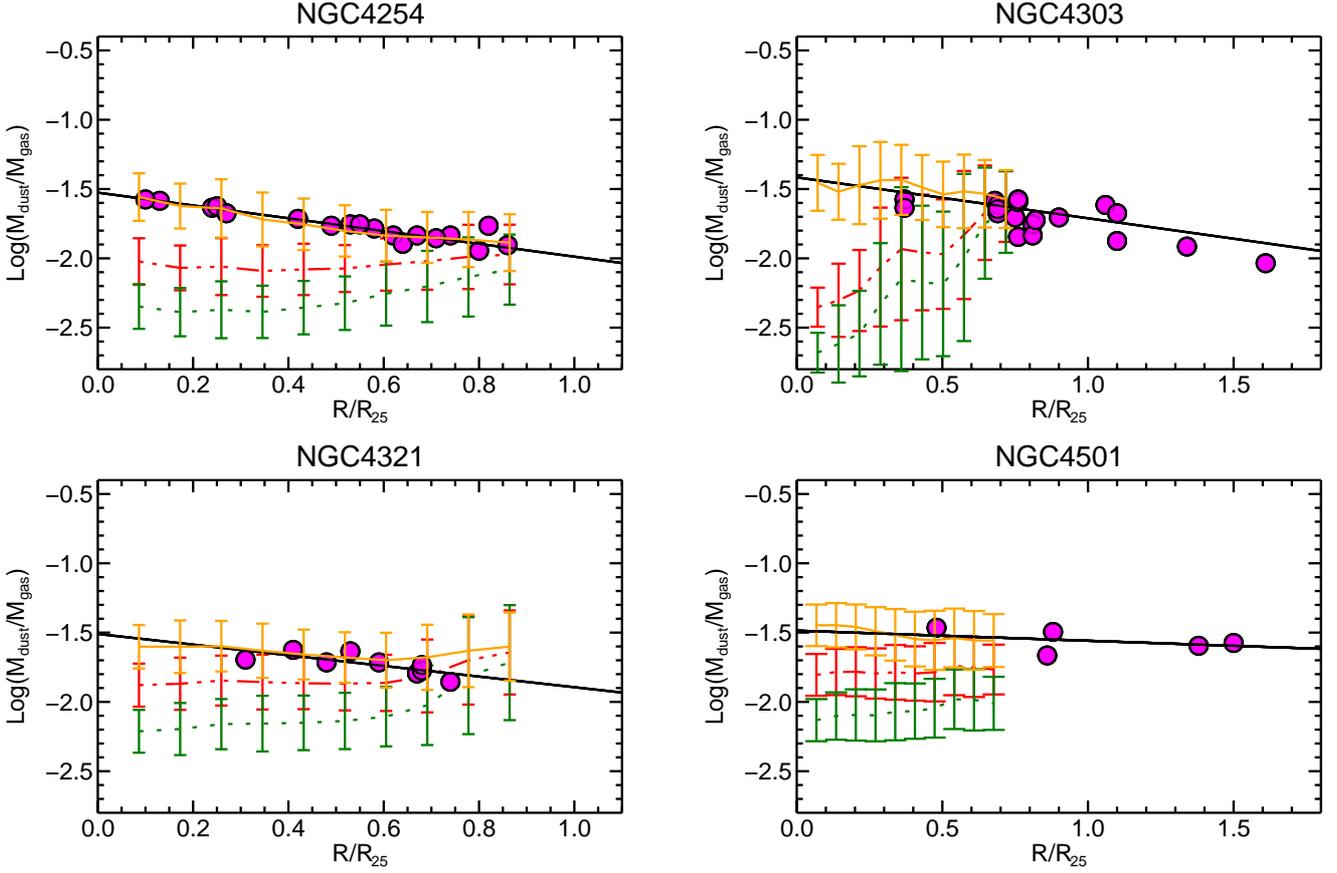}
\caption{Dust-to-gas mass ratio and metallicity radial profiles with the conversion factor  \xco\ 
0.1-0.9$\times10^{20}$ cm$^{-2}$ (K km s$^{-1})^{-1}$ (see text for the \xco adopted for each galaxy--orange solid  curves), and with constant   \xco=1.8$\times10^{20}$ cm$^{-2}$ (red dot-dashed  curves) and  \xco=4.0$\times10^{20}$ cm$^{-2}$ (green dotted curves). 
The magenta circles are the oxygen abundance converted to the dust-to-gas scale with Eq.3.  
The continuous black lines show the fit to the abundance data.  
 \label{newtest}}
\end{figure*}

 We have compared these gradients  also with the dust-to-gas mass ratio computed with  constant "Galactic" 
 \xco=1.8$\times10^{20}$ cm$^{-2}$ (Bolatto et al. 2008) and  \xco=4.0$\times10^{20}$ cm$^{-2}$ (Draine et al. 2007). 
The results are shown in Fig.\ref{newtest}:  with the low \xco, the negative slope of the metallicity gradient is reproduced by the dust-to-gas mass ratio gradient. 
We note also that the outer regions are marginally affected by the choice of \xco, and they are still consistent with a "standard" \xco\ values for normal disk galaxies, 
while the inner regions need a lower value to reproduce a decreasing dust-to-gas mass ratio. 
This is in agreement with the evidence found for a lower \xco\ in galaxy centers,  e.g., \citet{1994ApJ...428..638S}, \citet{2006A&A...445..907I}, \citet{2009A&A...506..689I},  \citet{2011MNRAS.411.1409W}.
On the other hand, the dust-to-gas mass ratio
obtained with    \xco=1.8$\times10^{20}$ cm$^{-2}$ and \xco=4.0$\times10^{20}$ cm$^{-2}$ tend to be flatter or positive, especially with the larger value of \xco. 
Thus, even if it is difficult to prove a dependence of \xco\ on the metallicity due to uncertainties on the O/H and dust mass scales, the comparison between the oxygen and  dust-to-mass mass ratio 
radial gradients in the four galaxies under analysis allows us to put a constraint on their \xco\ conversion factors, favoring lower values of \xco\ in the radial regions R$_G<$0.6-0.7R$_{25}$.

In Table~\ref{tab:h2} we report the total masses of HI, H$_{2}$, and dust. 
We computed the dust masses integrating the radial profile up to 0.7R$_{25}$. 
We remind that this limit is due to the requirement of a
S/N$>10$ at 500~$\mu$m, necessary 
to limit the uncertainties 
due to background subtraction and avoid the artefacts caused by the 
high-pass filtering in the PACS data reduction. 
The comparison 
with the results of \citet{davies11}, obtained with an integrated analysis of the galaxies, indicates that 
only 30\% of the dust mass resides in the outer-most regions, with R$>$R$_{25}$.


 \begin{table*}
\caption{HI, H$_{2}$, and dust masses}
\begin{tabular}{lllllll}
\hline\hline
Name 	& M(HI)                  & M(H$_2$)$_{\rm{unif.}}$  & M(H$_2$)$_{\rm{linear}}$&  M(H$_2$)$_{\rm{superlinear}}$ & M(dust)  \\
           	&10$^{9}$ M${\odot}$ &   10$^{9}$ M${\odot}$                                                  &10$^{9}$ M${\odot}$                                                 & 10$^{9}$ M${\odot}$  & 10$^{7}$ M${\odot}$    \\
    (1)       	& (2)                     & (3)     				          & (4) 		                        &(5)	            & (6) \\
\hline
NGC4254                	&  4.4$^{a}$  &  5.4 & 1.8  & 0.22 & 7.0\\
NGC4321	                  &  4.8$^{a}$    & 4.4  & 1.3  &0.13 & 8.6 \\
NGC4303		         &  3.3$^{b}$   & 4.8  & 0.8   & 0.06 &5.1  \\
NGC4501			&  2.7$^{a}$    & 3.8 & 1.0  & 0.05         & 7.4 \\
\hline
\hline
\label{tab:h2}
\end{tabular}
\tablefoot{ 
$a$ \cite{2009AJ....138.1741C} }\\
$b$ integration of the radial profile shown  in Fig.3 of \cite{1996ApJ...462..147S}.
\end{table*}

\citet{2010MNRAS.402.1409B} performed a similar analysis on NGC2403, 
a late-type galaxy with a relatively low molecular fraction. 
Because in NGC2403 the atomic gas is the largest component of the ISM, differing assumptions about 
\xco\ (they used a constant value and a metallicity dependence as in our Eq.~2) 
have only a minor impact on the radial profile of the dust-to-gas mass ratio. 
Unlike NGC2403, in the galaxies of our sample the molecular gas dominates the ISM 
within $\sim$0.5~R$_{25}$, and thus a variation of \xco\  has a significant effect
on the dust-to-gas ratio. 
\citet{2010MNRAS.402.1409B} found with both the constant and metallicity dependent \xco\ 
a good agreement between the O/H gradient 
(-0.084$\pm$0.009 dex kpc$^{-1}$) and the dust-to-gas gradient (-0.093$\pm$0.002 dex kpc$^{-1}$), 
with the dust-to-gas slope slightly steeper. 
They explain that  variations in the ratio of oxygen to the other constituents 
of dust might originate the difference in the gradients. 
For example, carbon and oxygen are both produced by supernovae, but carbon is also 
generated by the nucleosynthesis of low and 
intermediate mass stars; thus C/O is expected to vary radially,
as shown by  
\citet{garnett99}.  

\citet{2004MNRAS.351..362T} obtained a similar result 
in their study of the radial distribution of dust and gas in several nearby galaxies 
for which they had neutral hydrogen and 850~$\mu$m images available: 
their radial profiles of dust-to-gas mass ratio were flatter 
than the oxygen abundance gradients from HII regions.  
However in their galaxies CO observations were not available, so they were neglecting 
an important potentially dominant component in the inner regions.  

A similar result  was obtained also by \citet{2009ApJ...701.1965M} in the galaxies belonging to the SINGS sample. They have  studied the radial variation 
of the dust-to-gas ratio in a large sample of galaxies of different morphological types. 
They found that the dust-to-gas mass ratio decreases by an order of magnitude from the center to the edge of the optical disk of each galaxy, similarly to what happens for 
the gas-phase oxygen abundance. Their relation between dust-to-gas mass ratio and oxygen abundance can be described with a linear scaling law  for $12+\log$(O/H)$>$8.9, as we found 
for NGC4254, NGC4321, and NGC4501. At lower metallicities, they found that the dust-to-gas ratios are systematically below this simple relation. 
They explain this fact to be due to the  large amounts of gas that has not yet undergone star formation activity and that reside in the outer regions of spiral disks.

\subsection{Other evidence for a low \xco}

In agreement with our result, a value of 
\xco\ lower than $\sim$2$\times$ 10$^{20}$ cm$^{-2}$ (K km s$^{-1})^{-1}$ was
also expected in NGC4254 and NGC4321 from the analysis of \citet{2010A&A...518L..62E}.
They calibrated the mass-opacity coefficient ($\kappa_{\nu}$) using their dust-to-gas maps. 
Using a standard \xco\ value they found 
a mass-opacity coefficient significantly lower than the value (0.41  $\rm m^2\ kg^{-1}$) obtained by 
\citet{2002MNRAS.335..753J}.
To solve the discrepancy, they suggested that using only SPIRE bands they are 
over-estimating the temperature of the dust;  if the temperature of the 
dust in both NGC4254 and NGC4321 were actually $\simeq$10 K rather than 20 K, 
this would be enough to increase the value of $\kappa_{\nu}$ to that expected 
from the James et al. study. 
However, this is not the case since with our fit of PACS and SPIRE bands we found 
dust temperatures  in the range 17$<$ T $<$25 K found by \citet{2010A&A...518L..62E} (see Fig.~3).

The other possibility they suggested to reconcile their results with \citet{2002MNRAS.335..753J} 
was to have values of \xco$\simeq$6 times lower than the initial value they assumed, 
$\sim$2$\times$ 10$^{20}$ cm$^{-2}$ K km s$^{-1})^{-1}$.
Considering that we are using a  mass-opacity coefficient which is about half of 
that used by James et al., thus, to have agreement with our results, 
they would need \xco$\simeq$3 times lower than their standard value.  
This is consistent with our result; values of \xco\ in the range 
$\sim$0.5-1$\times$ 10$^{20}$ cm$^{-2}$ K km s$^{-1})^{-1}$
can reproduce both the correct value of the mass-opacity coefficient and the 
radial profiles of dust-to-gas mass ratio. 

\section{Summary and conclusions}

\begin{itemize}
\item[1.] We analyzed FIR observations obtained with PACS and SPIRE, 
together  with CO and HI maps from the literature, of four spiral galaxies in the Virgo cluster (NGC4254, 
NGC4303, NGC4321, and NGC4501). We derived the total mass of dust from SED fitting of the FIR images,
and used HI moment-1 maps to derive the 
geometric parameters of the galaxies, which are disturbed by tidal interactions. 
Finally, we extracted the radial profiles of atomic gas, molecular gas and dust, and compare them
with oxygen abundance radial gradients compiled from the literature and placed on a common 
abundance scale.  
\item[2.] To avoid the large uncertainties on the zero-points of the oxygen abundance 
derived with the bright-line methods, we used  the dust-to-gas ratio obtained with the 
Galactic value of  \xco\ to fix a lower limit to the oxygen abundance. 
This allowed us to discard a set of metallicity calibrations and to reduce the uncertainty 
on the oxygen gradient zero-point.  
We converted the literature O/H to the scale of KK04. 
\item[3.] We studied the dependence  of \xco\ on metallicity by  comparing the radial gradient of metal abundance in gas phase (oxygen abundance in HII regions)  
and in solid phase (dust in emission from the FIR versus the total gas--atomic and molecular), 
assuming that these two quantities decrease at the same rate with radius.
We considered a constant  \xco\ 
\citep[e.g.,][]{2008ApJ...686..948B},
\xco$\propto$Z$^{-1}$ ("linear'' dependence)
\citep[e.g.,][]{2002A&A...384...33B}, and
\xco$\propto$Z$^{-2.5}$ (``super-linear'' dependence)
\citep[e.g.,][]{2000mhs..conf..293I}.
A linear fit of  the gradient in the logarithm of the dust-to-gas ratio 
[d log(dust/gas)/dR$_{G}$] shows that,  within $\sim$0.7~R$_{25}$, 
 values of \xco\ in the range $\sim$0.5-0.9$\times$ 10$^{20}$ cm$^{-2}$ (K km s$^{-1})^{-1}$
were able to reproduce negative gradients, similar to the O/H ones for NGC4254, NGC4321, and 
NGC4501, while NGC4303 needs an extremely low \xco value   $\sim$0.1$\times$ 10$^{20}$ cm$^{-2}$ (K km s$^{-1})^{-1}$.
For NGC4254, these \xco\ values can be obtained with a linear  metallicity dependence and 
oxygen abundance calibrated by KK04. 
For NGC4303 a superlinear dependence is instead necessary to match within the errors the 
two gradients, while for NGC4501 and NGC4321 a \xco\ intermediate between the Galactic one and 
the linear metallicity dependence would be necessary (always with KK04 O/H calibration).
We suggest that a \xco\ lower than the standard Galactic one 
\citep[e.g.,][]{2008ApJ...686..948B} is necessary in these galaxies to obtain a decreasing 
dust-to-gas mass ratio similar to the O/H gradient of HII regions. 
These low \xco\ values are favored in the radial regions R$_G<$0.6-0.7R$_{25}$
where the molecular gas dominates, while  the outer regions, 
where the atomic gas is the main component, are less affected by the choice of \xco, 
and thus we cannot put constraints on its value. 
A large sample of galaxies with available  metallicity ad dust-to-gas mass ratio gradients 
is necessary to confirm and strengthen these results.  
\end{itemize}

\begin{acknowledgements}
We thank an anonymous referee for comments and suggestions which have improved the quality of the manuscript and its presentation. 
This research has made use of the NASA/IPAC Extragalactic Database (NED) which is operated by the Jet Propulsion Laboratory, California Institute of Technology, under contract with the National Aeronautics and Space Administration.
L.M. and C.P. are supported through an ASI-INAF grant "HeViCS: the Herschel Virgo Cluster Survey" I/009/10/0. 
C.V. received support from the ALMA-CONICYT Fund for the
Development of Chilean Astronomy (Project 31090013) and from the
Center of Excellence in Astrophysics and Associated Technologies (PBF
06).
\end{acknowledgements}

\end{document}